\documentstyle[aps,pre,array,amsmath,graphicx,multicol]{revtex}

\newcommand{\sP}{{\mathcal{P}}}
\newcommand{\Ps}{{\sP_{\rm shape}}}
\newcommand{\Py}{{\sP_{\rm yaw}}}
\renewcommand{\Pr}{{\sP_{\rm roll}}}
\newcommand{\Pt}{{\sP_{\rm tilt}}}

\begin{document}
\draft

\tightenlines

\newcommand{\bomega}{\mbox{\boldmath{$\omega$}}}

\title{Stability of Monomer-Dimer Piles}
\author{Deniz Erta{\c s}$^{(1)}$,  
Thomas C. Halsey$^{(1)}$, Alex J. Levine$^{(2)}$ and Thomas G. Mason$^{(1)}$}
\address{$^{(1)}$ Corporate Strategic Research, 
ExxonMobil Research and Engineering, Route 22 East, Annandale, 
New Jersey 08801 \\
$^{(2)}$ Department of Chemical Engineering, 
University of California, Santa Barbara,
California 93106}
\date{\today}
\maketitle

\begin{abstract}
We measure how strong, localized contact adhesion between grains affects 
the maximum static critical angle, $\theta_c$, of a dry sand pile. 
By mixing dimer grains, each consisting of two spheres that have been 
rigidly bonded together, with simple spherical monomer grains, we create 
sandpiles that contain strong localized adhesion between a given particle 
and at most one of its neighbors.  We find that $\tan\theta_c$ increases
from 0.45 to 1.1 and the grain packing fraction, $\Phi$, decreases from 
0.58 to 0.52 as we increase the relative number fraction of dimer particles 
in the pile, $\nu_d$, from 0 to 1. We attribute the increase in 
$\tan\theta_c(\nu_d)$ to the enhanced stability of dimers on 
the surface, which reduces the density of monomers that need to be
accomodated in the most stable surface traps. A full characterization
and geometrical stability analysis of surface traps provides a good 
quantitative agreement between experiment and theory over a wide range 
of $\nu_d$, without any fitting parameters. 
\end{abstract}

\pacs{45.70.Cc,61.43.Gt}

\begin{multicols}{2}

\section{Introduction}
The presence of adhesive forces between grains can greatly alter the 
physical behavior of sandpiles\cite{background1,background2}.  
  Although the importance of intergrain adhesion has been noted 
in many fields, ranging from soil science to civil engineering, the 
understanding of the physical principles governing the link between the 
macroscopic behavior of sandpiles in which adhesion is present and the 
microscopic attractive force distribution within the sandpile remains 
limited. 

Some light has been shed on this subject recently through 
experimental and theoretical studies which have shown that small quantities 
of liquid added to a sandpile comprised of rough spherical grains can cause 
sufficient intergrain adhesion so 
that the angle of repose after failure \cite{Hornbaker} and also the maximum 
static angle of stability of the sandpile before failure \cite{Mason}, 
known as the critical angle, $\theta_c$, greatly increase.  
A continuum theory that links 
stress criteria for the macroscopic failure of the wet pile to the 
cohesion between grains, which in turn was attributed to the formation
of liquid menisci with radii of curvature that were determined by the 
surface roughness characteristics of individual grains, 
has provided a satisfying quantitative explanation of the increase of
$\theta_c$ with the liquid volume fraction and air-liquid interfacial 
tension \cite{HL,Mason}. 

In this theory, each intergrain meniscus is assumed to exert the same 
average attractive force everywhere in the pile. This
assumption appears to be valid for sufficiently wet sandpiles;
however, when the volume of the wetting fluid is small enough this
theory is incapable of explaining the data.  There are two
principal features of the small--fluid--volume data that are
incompatible with the continuum theory\cite{HL}. The first is that the
increase in $\theta_c$ with small amounts of wetting fluid is
independent of the surface tension of that fluid\cite{Mason,Tegzes}.  
The second
is that, in order to quantitatively fit the data, one must assume
that a small fraction of the wetting fluid is sequestered on the
grains in such a way that it does not participate in the formation
of inter-grain menisci that contribute to the adhesive stresses
within the pile. It is possible that, at
vanishingly small fluid coverage, the physical/chemical
inhomogeneities of the grains' surface prevents the transport of
the wetting fluid from one mensicus to another thereby allowing a
wide distribution of inter-grain cohesive forces. To begin to
understand the effect of such a broad distribution of inter-grain
forces upon the macroscopic properties of the sandpile, we
consider in this paper an extreme example of such a distribution
in which some inter-grain contacts have an arbitrarily large
cohesion while others have no cohesion at all. Consistent with the
notion that the non-uniform distribution of inter-grain cohesion
is primarily significant at low fluid volumes, we study a system
designed so that each grain has at most one strong cohesive
contract.

In order to begin to investigate how nonuniform distributions of 
microscopic attractive forces between grains can affect the overall 
macroscopic stability of sandpiles, we have measured maximum stability
angles of dry sandpiles made by mixing a weight fraction $\nu_d$ of 
dimer grains (two spherical grains rigidly bonded together) into 
spherical monomer grains. 
The measured $\tan\theta_c(\nu_d)$ gradually increases 
over the entire range of $0<\nu_d<1$, 
despite a moderate drop in the total packing fraction of grains, 
$\Phi$, within the pile, due to the more inefficient packing of 
the dimers.  

A detailed theoretical study of the failure mechanism of such piles
leads us to the following key observations and conclusions, which enable 
us to quantitatively account for the increase in $\tan\theta_c(\nu_d)$:

(i) For piles consisting of perfectly rough particles 
(large intergrain friction), the stability of the free surface
upon tilting is limited by the particles on the surface layer,
which fail by rolling out of the surface traps they sit in.

(ii) For a given surface trap geometry, dimer particles typically
remain stable up to larger tilt angles; thus, for a mixture of monomers
and dimers, pile stability is limited by the monomers on the
surface, provided that the dimer concentration is not too large. 

(iii) Provided that individual grains rolling out of unstable
traps do not initiate avalanches, the pile will remain stable
as long as the density of {\it stable} surface traps is larger 
than the density of monomers on the surface layer.

(iv) The ratio of the density of monomers on the surface layer to
the total density of surface traps is $(1-\nu_d)/2$.

(v) A statistical characterization of the particle-scale roughness 
of the surface associated with grain packing is necessary to determine
$\theta_c$ quantitatively.  

The stability criteria for a pile consisting of a mixture of monomers and 
dimers with ideally rough surfaces can be cast as a purely geometrical
problem under conditions where rolling-initiated surface failure is the
primary mechanism that limits the stability of the pile. In this picture,
monomers and dimers on the surface layer occupy surface traps formed
by the particles underneath, each of which have a different stability
criterion associated with the trap's shape and orientation with respect to
the average surface normal and the downhill direction. For a pile to be 
stable at a given tilt angle, all the grains on the surface have to be 
sitting in a stable surface trap, suggesting that the least 
stable surface traps would control the overall stability of the pile. 

For a pile consisting of monomers, there are actually twice 
as many surface traps as surface grains,
and upon very gradually increasing the average tilt of the pile without 
disturbing the underlying grains, one finds that surface grains in  
traps that become unstable upon tilting 
can briefly roll down the pile's surface until they encounter an unoccupied 
stable surface trap which ends their descent, provided 
that at least half the surface traps remain stable so that all surface 
grains can be accomodated. 

A detailed analysis in Sec.~\ref{seccps} shows that it is much easier to 
trap dimers than monomers on a perfectly ordered close packed surface of 
spheres. This suggests that the stability of a random mixed pile of monomer 
and dimers is actually limited by the monomers on the surface, which 
accumulate in the most stable surface traps as $\theta_c$ is approached. 
If one assumes that the dimers remain essentially stable, one might expect 
that only a fraction $(1-\nu_d)/2$ of the surface traps remain 
stable when the pile is tilted to its maximum stability angle
$\theta_c(\nu_d)$; at this angle, there are just enough to accomodate all 
the monomers on the surface layer. A detailed experimental characterization 
of the positions of surface grains of random monomer and dimer piles
in Sec.~\ref{secanalysis}, combined with the computed stability 
criteria for monomers and dimers occupying surface traps in 
Sec.~\ref{secstability}, provide a quantitative 
explanation of the measured increase in $\theta_c(\nu_d)$ with no 
adjustable parameters up to 
$\nu_d \approx 0.6$, where the assumption of monomer failure begins 
to break down. 

This good agreement over a wide range of $\nu_d < 0.6$ 
reflects the subtle interplay between the distribution of orientational
and size fluctuations of the surface traps, some of which are stabilizing
and some destabilizing. Thus, accurate characterization of
``grain-scale" roughness (as distinct from the microscopic
surface roughness of the grains) is essential to achieve quantitative
agreement between theory and experiment. The important role played by
grain-scale roughness is also evident in the rheology of gravity-driven 
chute flows\cite{Ertas,Silbert}, where the precise nature of the bottom 
surface has significant influence on the resulting flow. Nevertheless, the 
results convincingly demonstrate that the stability of cohesionless
grains with large inter-grain friction is indeed controlled by 
surface failure, and should otherwise be insensitive to the type of grain
material.  
 
In addition to clarifying the role of surface failure in the stability
of piles by connecting macroscopic measurements of stability angle to
grain-scale composition, these results for a well-characterized 
sandpile also provide a critical link between the attractive forces 
within the sandpile and the nonspherical geometry of a well-known fraction 
of constituent grains.  In this respect, these measurements provide 
quantitative insight into similar measurements of the angle of 
repose after dynamic failure of less well-controlled piles of spheres 
and cylinders (e.g. peas and rice).  From this perspective, dimers may be 
imagined as short elongated grains that have surface irregularities of
the same order as the grain size. These irregularities promote the 
strong interlocking of adjacent grains, which inhibits the 
failure of the pile more than the typical contacts between smooth 
cylinders and ellipsoids.  

The rest of the manuscript is arranged in the following way. In 
Sec.~\ref{secexperiment}, we present the experimental method for preparing 
the dimer and monomer grains, measuring the critical angle of stability 
and grain packing fraction of the pile as a function of the dimer content, 
and characterizing the surface configuration of monomer and dimer piles. 
The results of these measurements are also reported in this section. The 
development of the theoretical understanding of these results starts in 
the next section, Section~\ref{secproblem}, in which the stability of a 
pile [of spherical grains] is posed as a geometrical problem. This section 
summarizes an earlier attempt to treat pile stability 
geometrically\cite{Barabasi} and presents a different and more 
general strategy for the solution, which can be extended to include dimer 
grains. Using this approach, in Sec.~\ref{seccps}, we fully solve the 
stability problem on a triangular close-packed surface layer for monomers 
and dimers. Section~\ref{secfluct} presents a statistical analysis of the 
measured shape and orientation of traps on real surfaces of piles 
comprised solely of either monomers or dimers, and uses the method 
outlined in Sec.~\ref{secproblem} to determine the corresponding 
solution to the stability angle. In Sec.~\ref{secconc}, we summarize the
main findings and insight gained from this study, as well as possible
future directions. 

\section{Experimental}
\label{secexperiment}
We prepare the dimer grains by bonding glass spheres of radius 
$d = 4.6\pm0.2$ mm and density $\rho_g =2.3$ g/cm$^3$ together 
using methyl acrylate glue.  The glue in its carrier 
solvent completely coats the surfaces of the glass spheres and accumulates 
in a contact meniscus between the spheres. After less than one day of 
drying, a strong shear-rigid bond between the spheres is formed.  The 
volume of the dried glue is much smaller than the volume of the spheres, 
so the two bonded grains have the appearance of an ideal dimer or doublet.  
Because the glue coats the entire surfaces of the grains and may thus alter 
the friction coefficient of the grains, we have likewise coated all the 
monomer grains with methyl acrylate so that the friction coefficient at the 
contact points between grains, whether monomers or dimers, is identical.  
The sandpiles are prepared by mixing together varying weight fractions 
$\nu_d$ of dimer grains into monomers, as determined by 
a balance. The mixtures are placed in a clear plastic box having a square 
bottom that is 6.5 cm wide and a height of 5.5 cm, yielding an average 
number of spheres per pile of about three hundred. 

\begin{figure}
\centerline{\includegraphics[width=3in]{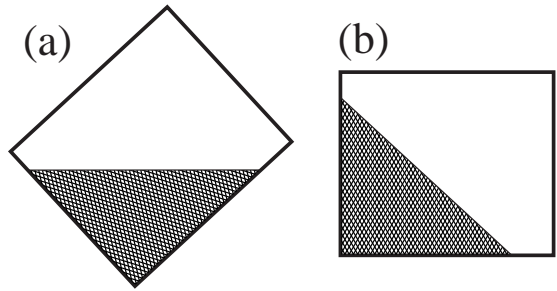}}
\caption{Measurement of the critical angle of stability, $\theta_c$.}
\label{figpile}
\end{figure}

To measure the critical angle, we employ a procedure that is identical to 
one that was used to study the critical angles of wet sandpiles\cite{Mason}. 
We tilt the box at an angle, as shown in  Fig.~\ref{figpile}(a), and shake 
it back and forth about five times along the direction of the lower edge
(normal to the page in the figure).  
This distributes the grains so that the surface of the pile is normal to 
the direction of gravity, as shown in Fig.~\ref{figpile}(b). There is no 
noticeable size segregation of the grains introduced by the shaking. We 
have purposefully avoided tapping the pile in order to prevent a 
densification that could affect the pile's surface characteristics and 
stability\cite{UChicago}. We then place the lower 
edge of the box on a table and slowly tilt it so that the bottom rests 
flat on the table and normal to the direction of gravity. If the pile fails 
catastrophically, then no measurement is recorded, but if the pile remains 
stable after only isolated movement and resettling of a few grains, then a 
measurement of the static angle of the pile, $\theta$, is recorded, as 
shown in Fig.~\ref{figpile}(c). After several trials, a rough determination 
of the stable angle is obtained, and thereafter the initial angles are not 
chosen randomly, but instead are kept close to this value. Using the 
results of ten trials, we average the values of the three largest angles 
to obtain the critical angle, $\theta_c$.  This value of the critical 
angle is reproducible; the variation in the three angles used to obtain 
the average is about ten percent for all values of $\nu_d$. 
This procedure yields slightly larger angles than those found in typical 
angle of repose measurements in which the sandpile is induced to fail.

\begin{figure}
\centerline{\includegraphics[width=3in]{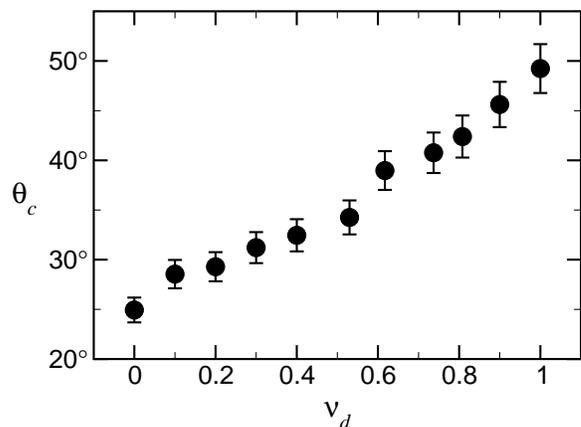}}
\caption{The critical angle of stability, $\theta_c$, as a function of 
dimer weight fraction, $\nu_d$.}
\label{figthetac}
\end{figure}

Figure \ref{figthetac} depicts $\theta_c$ as a function of $\nu_d$.  
We find that it increases approximately linearly,
from $\tan\theta_c \approx 0.45$, a well established value for a wide 
variety of dry spherical grains\cite{Barabasi}, to 
$\tan\theta_c \approx 1.1$ for a sandpile 
comprised completely of dimers. 

We have also measured the average packing 
volume fraction of grains, $\Phi$, in the sandpile as a function of 
$\nu_d$ by measuring the mass of water required to fill the voids
in the 
pile as it stands in the tilted configuration shown in Fig.~\ref{figpile}(a).
Based on the relative number of grains at the surfaces compared to those 
within the pile, we estimate that the lowering of the local grain packing 
fraction due to the open or wall surfaces makes the measured $\Phi$ 

\begin{figure}
\centerline{\includegraphics[width=3in]{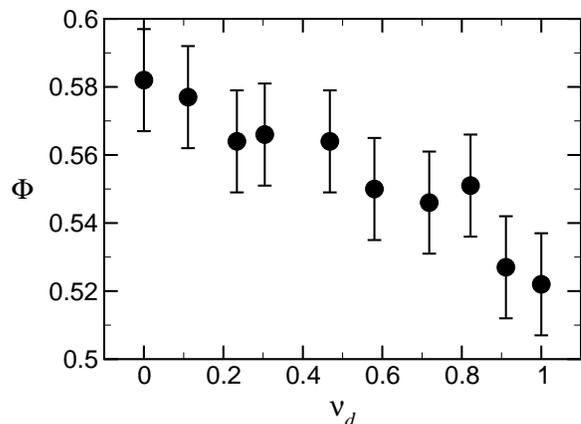}}
\caption{The packing fraction, $\Phi$, as a function of 
dimer weight fraction, $\nu_d$.}
\label{figdens}
\end{figure}

\noindent  appear 
to be about two percent smaller than the corresponding bulk value. These 
measurements of $\Phi(\nu_d)$ are
plotted in Fig.~\ref{figdens}. 
The overall reduction in the packing fraction of about ten percent
indicates that dimer grains pack less efficiently than monomer grains.

In order to experimentally characterize the surfaces of random piles of 
either monomer or dimer grains, we have taken stereo digital images of 
stable piles using top and front views in order to reconstruct the 
three-dimensional coordinates of the centers of all of the spheres 
visible on the surfaces of the piles. We do not include spheres that 
touch walls or the bottom surface of the container. Due to the significant 
surface roughness of the piles, especially in the case of the dimer pile, 
we occasionally detect the position of a sphere that lies more than one 
diameter below the average surface defined by all the grains. 
These spheres are not true surface spheres and are eliminated from 
consideration in the subsequent surface trap analysis (See 
Sec.~\ref{secanalysis}.) 

\begin{figure}
\centerline{\includegraphics[width=3in]{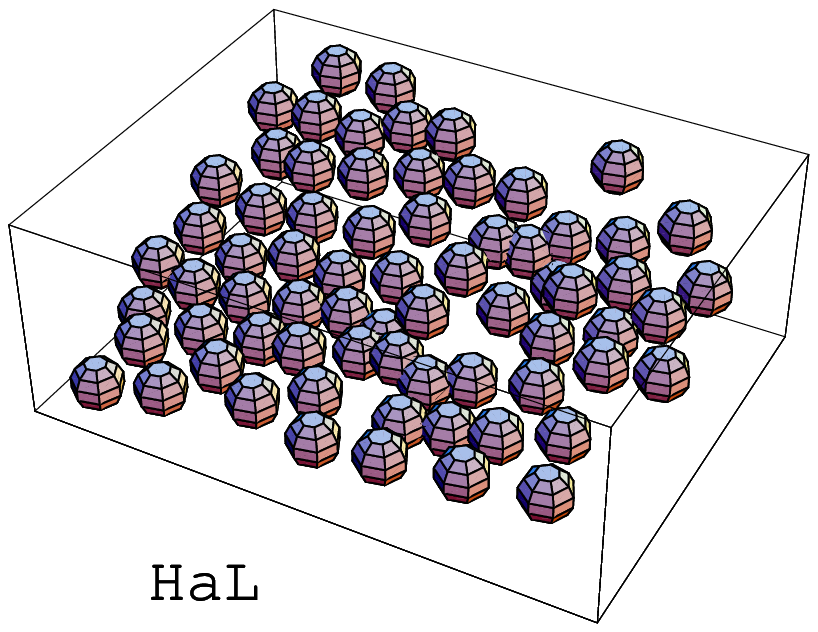}
\includegraphics[width=3in]{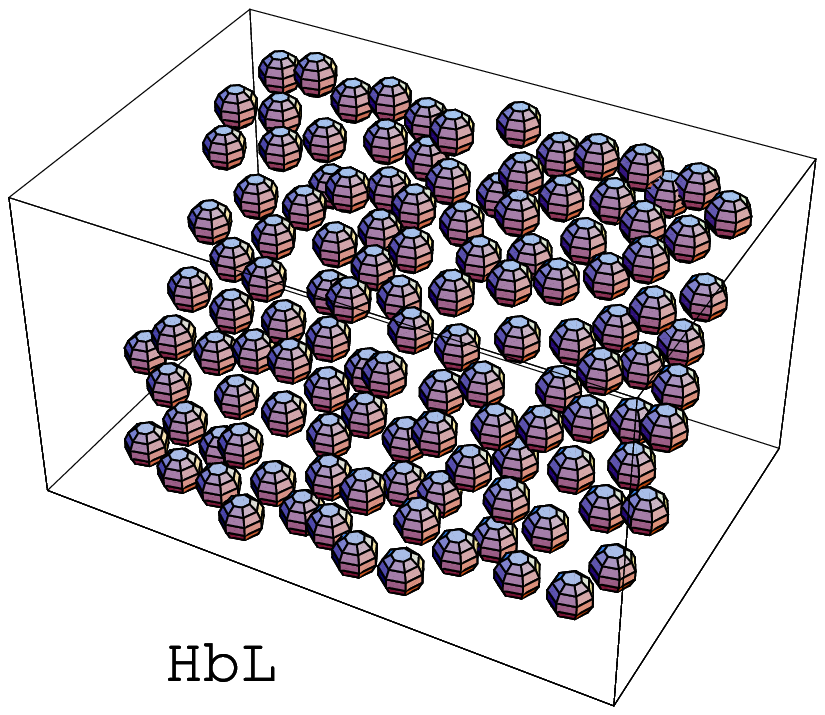}}
\caption{Reconstructed positions of spheres at the surface of a pile 
consisting of (a) monomers, and (b) dimers. The technique used can
identify the spheres, but not the bonds of the dimers, thus only half
of a dimer might be shown in some cases. However, that information is not 
needed in the subsequent analysis in Sec.~\protect\ref{secanalysis}.}
\label{figpiles}
\end{figure}

The measured positions of the surface spheres for 
a stable monomer pile and a dimer pile are shown in Fig.~\ref{figpiles}(a) 
and \ref{figpiles}(b), respectively. The monomer pile shown is very
close to its critical angle of stability $(\theta=23^\circ)$, whereas 
the dimer pile shown here, while at a much higher angle ($\theta=41^\circ$), 
is still somewhat below its $\theta_c$. Finally, we have 
qualitatively observed that the roughness of the sandpile's free surface 
increases somewhat as more dimers are included in the pile. This increase 
has been quantified in terms of larger fluctuations in the shapes and 
orientations of the surface traps, as presented in Sec.~\ref{secanalysis}.

\end{multicols}

\begin{multicols}{2}

\section{Pile stability as a geometrical problem}

\label{secproblem}

\subsection{Background and Context}

\begin{figure}
\centerline{\includegraphics[width=3in]{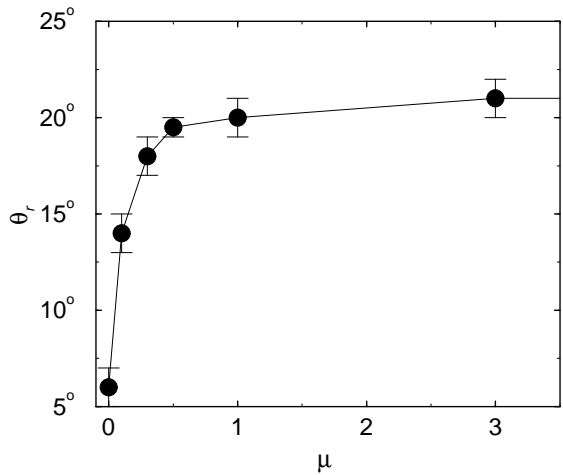}}
\caption{Angle of repose for a pile of spheres with Hertzian 
contact interactions and static friction as a function of 
friction coefficient $\mu$ betweeen the spheres, 
obtained from granular dynamics simulations by extrapolating 
flow rates as a function of tilt angle to zero flow. Details 
of the simulation technique can be found in 
Refs.\protect\cite{Ertas,Silbert}.
}
\label{figthetavsmu}
\end{figure}

Various measurements of the angle of repose $\theta_r$ 
for cohesionless piles of smooth spherical particles
come up with the same value of about 22$^\circ$, largely 
independent of the makeup of the spheres or of their surface 
properties\cite{Barabasi}. On the other hand, 
the shape of particles in a pile has a large influence 
on $\theta_r$ and the slightly larger critical angle 
 $\theta_c$\cite{ricepiles}. Furthermore, our granular 
dynamics simulations of piles, made of spheres with Hertzian 
contacts and static friction, show that $\theta_r$ initially 
increases rapidly with increasing friction coefficient $\mu$, 
although it saturates at a value of about $22^\circ$ for 
$\mu > 1$ (see Fig.~\ref{figthetavsmu}).

These observations suggest a primarily geometrical origin for the 
robustness of $\theta_c$ for perfectly rough spheres ($\mu \gg 1$), 
which can be further studied in an idealized system in which sliding 
is disallowed, due either to a very large  friction coefficient or 
to interlocking surface irregularities. The spheres in such 
a static pile can be classified into two groups as follows: 
A surface layer that
consists of spheres held in place by exactly three spheres
and their own weight, and interior spheres that have more than three 
contacts. (We do not consider the small population of ``rattler" 
spheres with three contacts that might exist in the interior of the 
pile, which will presumably not influence the stability of the pile.)  
The centers of the three spheres that support each sphere in the
surface layer form the vertices of what we henceforth call the ``base
triangle" associated with that sphere. 

As the tilt angle of the pile is increased, spheres on the 
surface layer can move by rolling out of the surface trap formed by 
the three supporting neighbors. However, interior particles (excluding
rattlers) are held in place by a cage formed by their contacting 
neighbors, and if the friction coefficient is sufficiently large 
to preclude any sliding, they cannot move until this cage is 
destroyed by the motion of at least one of their neighbors.
This suggests that initiation of failure occurs at the surface layer, 
provided that sliding is disallowed. If the coefficient for 
rolling friction is small enough to be neglected, the stability of 
a sphere on the surface layer, and consequently the 
determination of $\theta_c$, becomes a purely geometrical problem. 
For finite values of $\mu$, other failure mechanisms can be
expected to reduce $\theta_c$ from this surface controlled value,
as observed in Fig.~\ref{figthetavsmu}.

A recent attempt at a theoretical determination of 
$\theta_c$ from the perspective of surface 
stability was made by Albert and co-workers\cite{Barabasi}. 
They have considered the stability of spheres at the surface,
supported by three close-packed spheres that form a base triangle,
and calculated the tilt angle $\theta_{max}$ at which the sphere would 
roll out of the trap formed by the base triangle as a function 
of ``yaw" $\phi$, i.e., the relative angle of orientation of the triangle
with respect to the downslope direction:
\begin{eqnarray}
\label{eqsphere}
\tan\theta_{max}(\phi)&=&\frac{1}{2\sqrt{2}\cos(\phi)}, 
\quad |\phi| < \frac{\pi}{3}, \\
\theta_{max}(\phi)&=&\theta_{max}(\phi+2\pi/3).
\end{eqnarray}
This stability criterion is periodic with period $2\pi/3$ due to symmetry. 
Yaw $\phi=0$ corresponds to an orientation in which one of the edges of the
base triangle is perpendicular to the downslope direction.
In order to account for the randomness in the orientations
of the base triangles on a disordered surface, they have suggested that 
the appropriate value for $\theta_{c}$ can be obtained by averaging 
$\theta_{max}$ over yaw $\phi$; they assumed a uniform distribution for this 
quantity. This yielded a value for $\theta_c$ that closely matched 
experimental observations. 

\subsection{Approach}

As already pointed out by its authors, the calculation in 
Ref.~\cite{Barabasi} represents a mean-field approximation, since it 
ignores variations in the shape and orientation 
of individual surface traps, which can be parametrized by  
their local tilt, yaw and roll angles, and the actual edge lengths of 
the base triangles. (For definitions of the these parameters, 
see the Appendix.)
Nevertheless, their result is in good agreement 
with experiments. We will address the dependence of the stability 
of a surface trap on some of these additional parameters in more 
detail in Section\ref{secstability}.

There is, however, a more serious complication with the
averaging approach than the neglect of fluctuation effects: The
stability of the pile  requires {\it all} particles on 
the surface to be stable. Thus, the stability of the pile should
be dictated by the particle in the {\it least stable} surface
trap, and not an average stability criterion. One might thus
wonder why the averaging approach appears to work so well.

In fact, for a pile of monomers, the number of base triangles 
forming potential surface traps is essentially twice the number 
of surface particles that actually reside in them. This is easy 
to see in the case of close-packed layers, where each 
successive layer to be placed on top has a choice among two 
sublattice positions; this is what gives rise to random stacking. 

The relationship is actually more general. If the surface 
of a pile is sufficiently smooth such that an average surface normal 
vector to the pile can be determined (note that this is a prerequisite
to actually being able to define and measure $\theta_c$), the base 
triangles associated with surface traps can be identified 
by a Delaunay triangulation of the sphere centers at the surface layer,
projected onto the plane of the mean pile surface. In such a triangulation,
the number of triangles per surface layer sphere is exactly two, since
the sum of all the interior angles of the triangles is 
$\pi\times$(no. triangles)$=2\pi\times$(no. surface particles).
An intrinsic assumption here is that the surface layer is similar
to the ``sublayer", consisting of those spheres that would become
part of the new surface layer if all the original surface particles were
removed simultaneously. The triangulation procedure to identify the 
surface normal vector and all of the potential surface traps is discussed 
in greater detail in Sec.~\ref{secanalysis}.

This ratio of surface trap to surface sphere density indicates that
in a stable pile, only half of the traps are actually filled. 
The pile will then find a stable configuration
as long as at least half of the surface traps are stable at the given 
tilt angle of the pile, since surface spheres that are in unfavorable 
traps can roll down the slope until they find a vacant trap of 
sufficient stability, assuming that they do not gain enough kinetic 
energy to knock other particles off their traps and cause an 
avalanche. Continuous failure of surface spheres will 
occur if there are never enough traps to stabilize the entire layer.  

This leads to the conclusion that the stability of the pile is actually
determined by the {\it median} stability angle of the traps, not the
mean. Nevertheless, as shown in Sec.~\ref{seccps}, the quantitative 
difference between this criterion and that studied in 
Ref.~\cite{Barabasi} is small; about 1.6 degrees. 

\section{Monomer-dimer stability on a flat close-packed surface}
\label{seccps}
 
Before launching a full-scale analysis of the pile stability problem
for random piles having random surface grain configurations, it is 
instructive to consider the implications and power of this new approach 
to stability on a simplified system consisting of a mixture of spheres 
(monomers) and dimers sitting on a triangular close-packed lattice. The 
stability analysis leading to 
Eq.(\ref{eqsphere}) can be extended to dimers. Dimers sit in surface 
traps such that the vector connecting the centers of the two spheres 
forming the dimer are always parallel to one of the edges of the base 
triangles, thus their orientation with respect to the downhill direction 
can be described by the angle $\phi$ as well. The resulting stability 
angle as a function of $\phi$, defined in the interval $(-\pi,\pi)$, is:
\begin{equation}
\label{eqdimer}
\tan\theta_{max}^{\rm dimer}(\phi)=
\begin{cases}
\frac{1}{2\sqrt{2}\cos(\phi)}, & |\phi| < \arctan(3\sqrt{3}),\\
\frac{\sqrt{2}}{\cos(\phi-2\pi/3)}, & \arctan(3\sqrt{3}) < |\phi| < 2\pi/3,\\
\frac{1}{\sqrt{2}\cos(\phi-\pi)}, & 2\pi/3 < |\phi| < \pi.
\end{cases}
\end{equation}

\begin{figure}
\centerline{\includegraphics[width=3in]{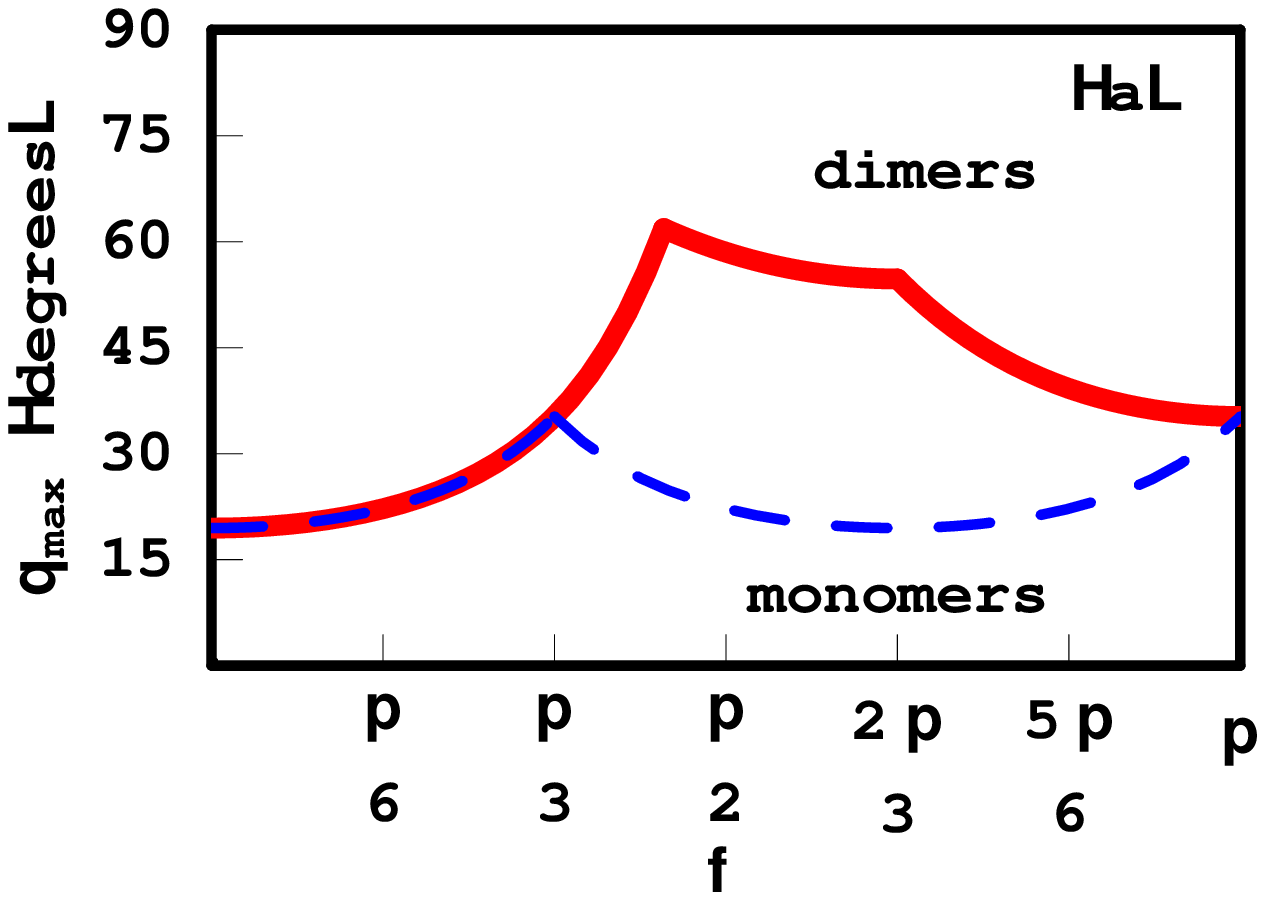}}
\medskip
\centerline{\includegraphics[width=3in]{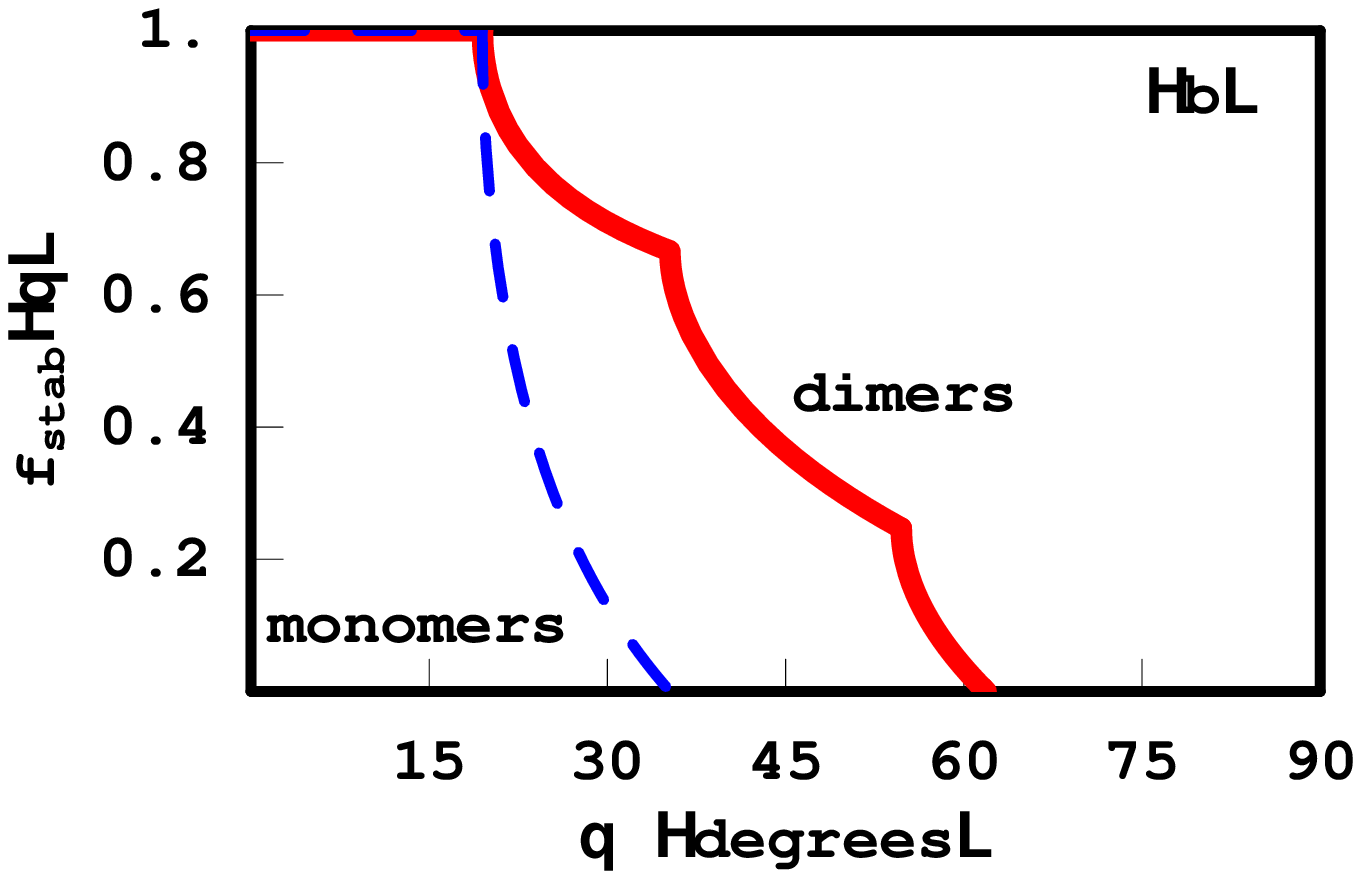}}
\medskip
\caption{(a) Stability angle of monomers (dashed line) and 
dimers (solid line) on a close-packed surface as a function of 
orientation (yaw) $\phi$. Monomer and dimer curves are
identical for $0<\phi<\pi/3$.
(b) The fraction of stable traps at a given tilt angle $\theta$
corresponding to a population of monomers (dashed line) and 
dimers (solid line) in surface traps with a uniform
yaw distribution.}
\label{figstab}
\end{figure}

The two functions defined by Eqs.(\ref{eqsphere}) and (\ref{eqdimer})
are plotted in Fig.~\ref{figstab}(a). It is striking how much more
stable dimers are compared to spheres for certain orientations
of the traps. This is due to the more favorable position of the
center of mass of the dimer, located where the two spheres meet,
which makes it more difficult to roll out of the surface traps.
An alternate, and perhaps more vivid, way of seeing the relative 
stability of dimers with respect to monomers is to plot the fraction
of stable traps $f_{\rm stab}(\theta)$ at a given tilt angle 
$\theta$:
\begin{eqnarray}
\label{eqfstab}
f_{\rm stab}(\theta)&\equiv&\int_{\theta}^{\pi/2} d\theta' DOS(\theta'), \\
\label{eqDOS}
DOS(\theta)&\equiv&\int_{-\pi}^{\pi}d\phi 
\Py(\phi)\delta\left(\theta-\theta_{max}(\phi)\right), \\
\label{eqpyaw1}
\Py(\phi)&=&\frac{1}{2\pi},\quad -\pi<\phi<\pi. \\
\end{eqnarray}

In the above, $f_{\rm stab}(\theta)$ is defined in terms of the  
density of surface traps at a given stability
angle, $DOS(\theta)$. 
For this particular case, the distribution of
yaw $\Py(\phi)$ is assumed to be uniform in the interval $(-\pi,\pi)$,
corresponding to an isotropic surface geometry. The resulting plot
of $f_{\rm stab}(\theta)$ for monomers and dimers is shown in 
Fig.~\ref{figstab}(b), and clearly demonstrates the difference in 
their stability. Consequently, the critical angles of stability 
inferred from the median stability angle for monomers and dimers are:
\begin{eqnarray}
\theta_{c}^{\rm monomer}&=&22.2^\circ, \\
\theta_{c}^{\rm dimer}&=&38.3^\circ.
\end{eqnarray}
Values obtained through the averaging procedure of 
Albert {\it et al.}\cite{Barabasi} are only slightly
larger: $\theta_{c}^{\rm monomer}=23.8^\circ$\cite{error} 
and $\theta_{c}^{\rm dimer}=39.5^\circ$.

For a surface layer consisting of a mixture of spheres and dimers, 
with a dimer volume fraction of $\nu_d$, the stability of the 
surface layer will be primarily controlled by the 
monomers, since dimers will be stable at most locations on the surface
at $\theta_c$ and do not need to be considered as surface particles
for the purposes of the stability analysis. Thus, for a given tilt angle, 
the surface layer can find a stable configuration as long as the fraction 
of stable traps at that angle, $f_{\rm stab}(\theta)$, exceeds
$(1-\nu_d)/2$, the density needed to accomodate all the monomers.

\begin{figure}
\centerline{\includegraphics[width=3in]{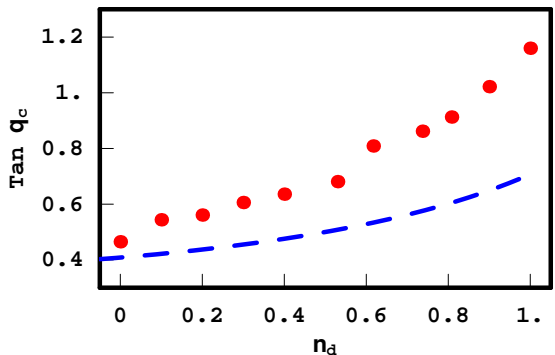}}
\caption{Measured and calculated values of $\theta_c$ as a function
of dimer weight fraction $\nu_d$. Circles: Experiment. Dashed Line:
Computation based on a randomly oriented, flat, {\it close-packed} surface.}
\label{fignodistcomp}
\end{figure}

The sequential filling of surface traps starting from the most stable 
one is somewhat analogous to the filling of an energy band in a 
fermionic system, with $-\theta_{max}$ for a trap corresponding to the 
energy $E$ of a fermionic state. $DOS(\theta)$ can then be interpreted as 
a ``Density of States". The monomer pile is analogous to a 
half-filled energy band and the addition of dimers lowers the filling 
fraction from 1/2. Thus, the critical stability angle 
$\theta_c^{\rm mix}(\nu_d)$ is determined by the ``Fermi energy" of the 
system at the given filling fraction, defined through the implicit
relation
\begin{equation}
f_{\rm stab}\left(\theta_{c}^{\rm mix}(\nu_d)\right)=\frac{1-\nu_d}{2}.
\label{eqtheta1}
\end{equation}
This relation has been plotted as a dashed line along with 
experimental data for $\theta_c$
in Fig.~\ref{fignodistcomp}. Although it captures the essential features
of the dependence on dimer mass fraction and provides a compelling
mechanism for this effect, the results are not quantitatively comparable.
The origin of the discrepancy lies primarily in the simplifications
made in characterizing the surface: Fluctuations in the shape and 
orientation of surface traps will broaden the DOS spectrum and 
consequently change the values obtained for $\theta_c$. 
In Section~\ref{secfluct}, we include the most relevant of such
fluctuations in the analysis and compare results to available experimental
data on the properties of such surfaces. Agreement between theory and
experiment improves significantly when such fluctuation effects are
taken into account.    
 
\section{Monomer Stability on a Random Surface}
\label{secfluct}

In generalizing the approach of the previous section to random
surfaces, we will assume that the stability of the pile is 
still controlled entirely by the monomers. This assumption is likely
to break down at very large dimer concentrations, so that comparison
to experiments may not be appropriate in that case. However,
it enables the determination of stability angles based entirely on
the behavior of monomers, and avoids the extremely tedious
analysis of dimer surface traps. Monomer traps, on the other hand,
are completely characterized by their base triangle, formed
by the centers of the three supporting spheres.
   
There are two steps that are needed to obtain the monomer DOS
required to compute $\theta_{c}^{\rm mix}(\nu_d)$ through
Eq.~(\ref{eqtheta1}). The first step is to determine the stability 
criterion for individual surface traps as a function of their 
shape and orientation. The second step is to develop an
adequate statistical description of the distributions of 
these surface traps as functions of the shape and orientation
parameters identified in the first step. 

\subsection{Stability of a surface trap}
\label{secstability}

For a surface trap of specified geometry, represented by its 
base triangle, what is the angle to which the pile can be 
tilted until the trap can no longer stably support a sphere 
and the sphere would roll out? In order to answer this question, 
we first need to quantitatively describe the geometry of the surface
trap with respect to the surface of the pile. This is done
in the Appendix, where the yaw $\phi$, roll $\psi$ 
and tilt $\theta$ of a base triangle are defined (See Fig.~\ref{figgeom}.)
 
The determination of stability criteria as a function of
shape, yaw and roll is a straightforward but tedious job.
We have used a Mathematica notebook to compute the stability
diagram for equilateral traps as a function of normalized
average edge length $a\equiv(l_1+l_2+l_3)/(3d)$, yaw $\phi$, and 
roll $\psi$, where $d$ is the diameter of the spheres. 

\begin{figure}
\centerline{\includegraphics[width=3.1in]{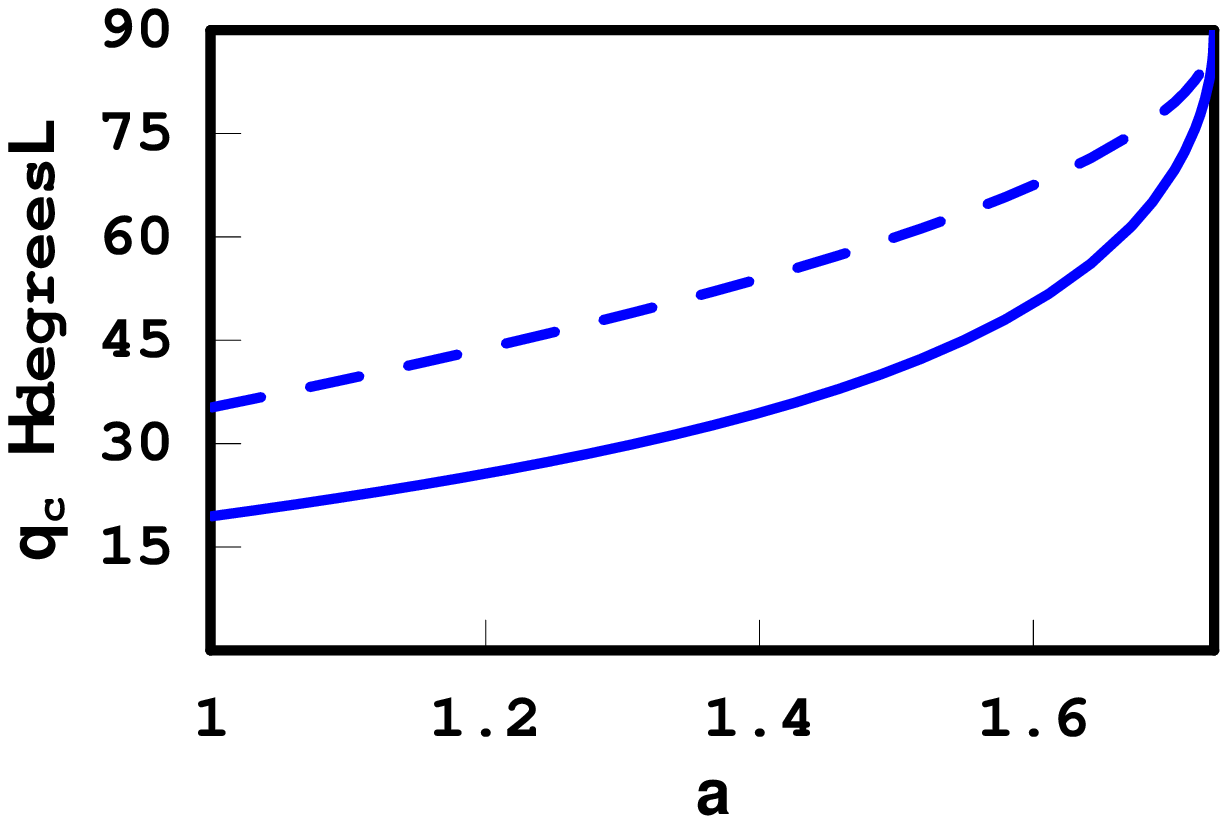}}
\centerline{\includegraphics[width=3.1in]{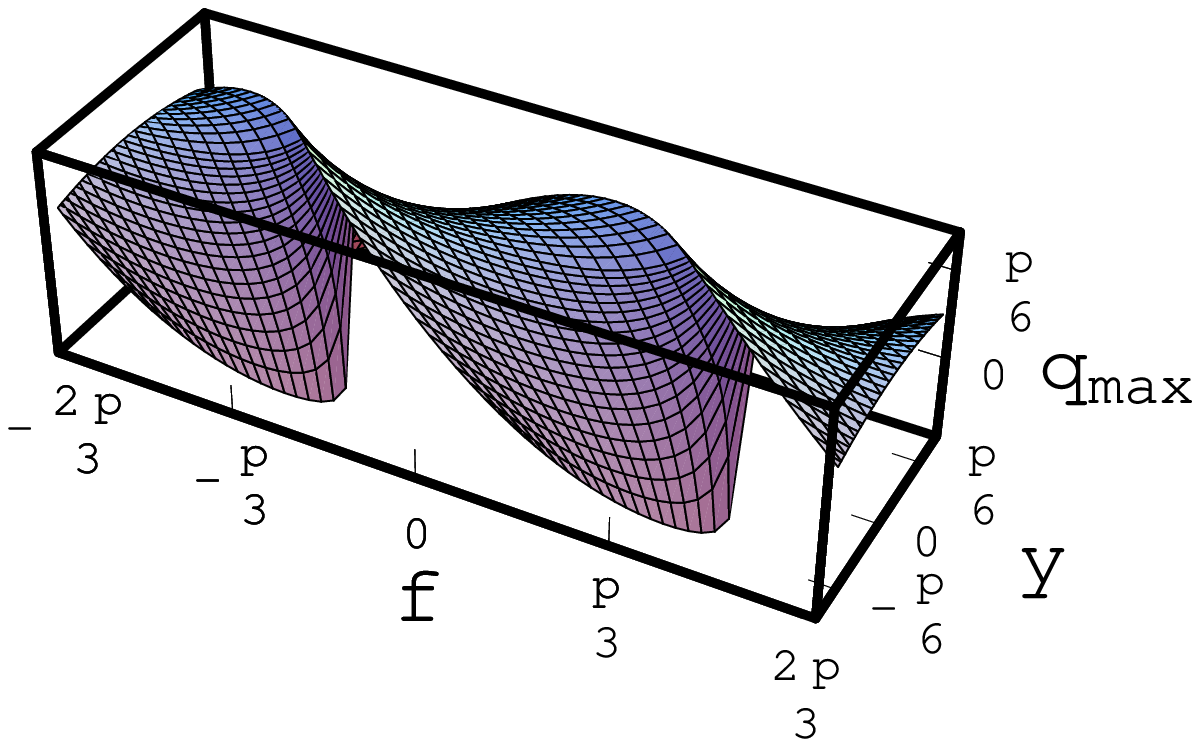}}
\caption{The maximum stability angle $\theta_{max}$ of equilateral
traps as a function of (a) normalized edge length $a$ for 
roll $\psi=0$, and yaw $\phi=0$ (solid line) and $\phi=30^\circ$ (dashed line); 
(b) $\theta_{max}$ as a function of yaw $\phi$ and roll $\psi$ for $a=1.13$. 
}
\label{figthetamax}
\end{figure}

The dependence of the maximum stability angle
$\theta_{max}$ on $a$ for traps with $\psi=0$ is plotted 
in Fig.~\ref{figthetamax}a. It is clear that this
parameter greatly influences the stability of the pile.
In order to estimate the value of $a$ for a random packing
of spheres with a given packing fraction $\Phi$, let us 
consider the tetrahedra in a Delaunay tessellation of 
the packing. Provided that the number of
tetrahedra per sphere do not change for the packings
of interest, the average volume of the tetrahedra
will vary as $V_{\rm tet}\sim\Phi^{-1}$. Spheres on the surface layer
will settle into the minima of their traps, thus the
tetrahedra they from together with their three supporting
spheres always have three edges whose lengths are 
equal to the diameter $d$. Thus, the average edge length 
of the faces that form the base triangle is expected to
vary as $l_{av}\sim\Phi^{-1/2}$. Since all edge lengths 
are equal to $d$ for the densest packing 
with $\Phi_{cp}=0.74$, the estimate for the average 
normalized edge length of base triangles is
\begin{equation}
a(\Phi)\approx\left(\frac{\Phi}{0.74}\right)^{-1/2}.
\label{eqa}
\end{equation} 
For the monomer pile with $\Phi=0.58$, this gives
$a=1.13$, in agreement with direct measurements
done on the pile (see Sec.~\ref{secanalysis}.)
  
Fig.~\ref{figthetamax}b shows 
$\theta_{max}$ against $\phi$ and $\psi$ for surface traps 
with $a=1.13$. (For certain values of yaw and roll, there 
is no tilt angle for which the traps are stable, and therefore 
$\theta_{max}$ is undefined.)

The analysis in Ref.\cite{Barabasi} is more limited in the 
types of surface traps it considers, as it only looks at
traps with $a=1$ and $\phi=0$. As seen in 
Fig.~\ref{figstability}, roll and edge length have great 
potential impact on the stability of a surface trap.
Although the Mathematica notebook can determine the stability 
diagram for the most general case, we will restrict our 
analysis to equilateral traps in order to keep the subsequent 
analysis tractable.

\subsection{Statistical description of traps on the surface of
a pile}
\label{secanalysis}
Having characterized and obtained stability criteria for a
given surface trap, the next task is to obtain a statistical
description of their population, through probability 
density functions (PDFs). In this study, we will neglect
short-range correlations between adjacent traps, e.g., 
associated with the sharing of edges, and assume that they
are drawn independently from an ensemble described by
PDFs for the values of edge lengths, yaw, roll, and tilt.
For the present, we will assume that all traps are equilateral
triangles with edge length $ad$, with uniformly distributed yaw 
angles and Gaussian roll and tilt angle distributions:
\begin{eqnarray}
\label{eqps}
\Ps(l_1,l_2,l_3)&=&\delta(l_1-ad)
\delta(l_2-ad)\delta(l_3-ad), \\
\label{eqpyaw}
\Py(\phi)&=&\frac{1}{2\pi},\quad -\pi<\phi<\pi, \\
\label{eqproll}
\Pr(\psi)&=&\frac{1}{2\pi\sigma_\psi^2}e^{-%
\frac{\psi^2}{2\sigma_\psi^2}}, 
\end{eqnarray}
\begin{eqnarray}
\label{eqptilt}
\Pt(\theta)&=&\frac{1}{2\pi\sigma_\theta^2}e^{-%
\frac{(\theta-\theta_{\rm pile})^2}{2\sigma_\theta^2}}.
\end{eqnarray}
In the above, $\{l_i\}$ correspond to the edge lengths;
we will neglect the variability in the shapes of the traps
and focus on equilateral traps of uniform size in order 
to study the effect of orientational disorder. 
If desired,
the subsequent analysis can be generalized to study the 
impact of disorder in the shapes of the traps as well.

The orientational PDFs are motivated by assuming that the pile 
surface was created with no initial tilt, and rotationally 
isotropic in the plane of the surface, and that little or no 
rearrangement took place in the surface traps during the 
subsequent tilting of the pile. This would result in a uniform
PDF of yaw, and nearly identical PDFs for roll and tilt 
($\sigma_\psi\approx\sigma_\theta$)\cite{commutation}. 

\begin{figure}
\centerline{\includegraphics[width=3.1in]{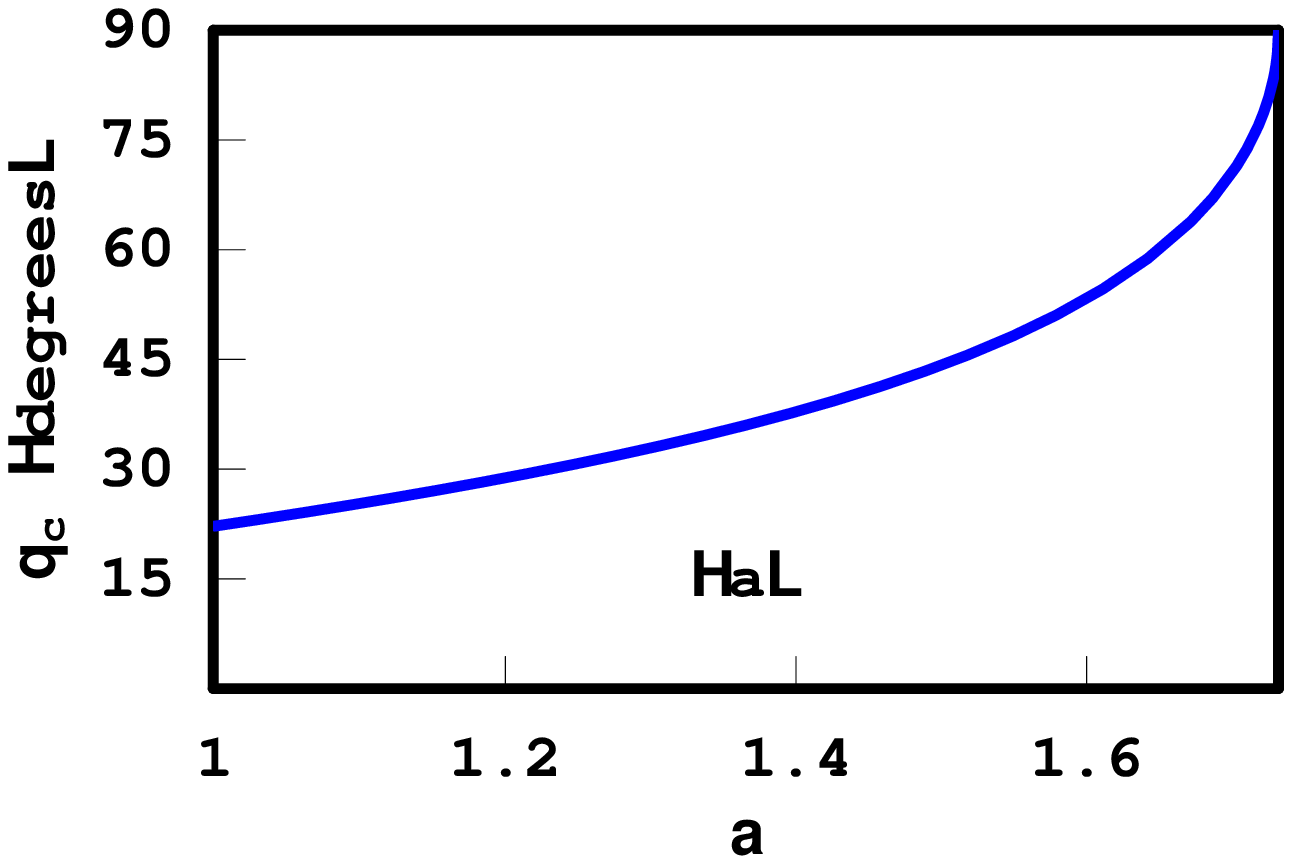}}
\centerline{\includegraphics[width=3.1in]{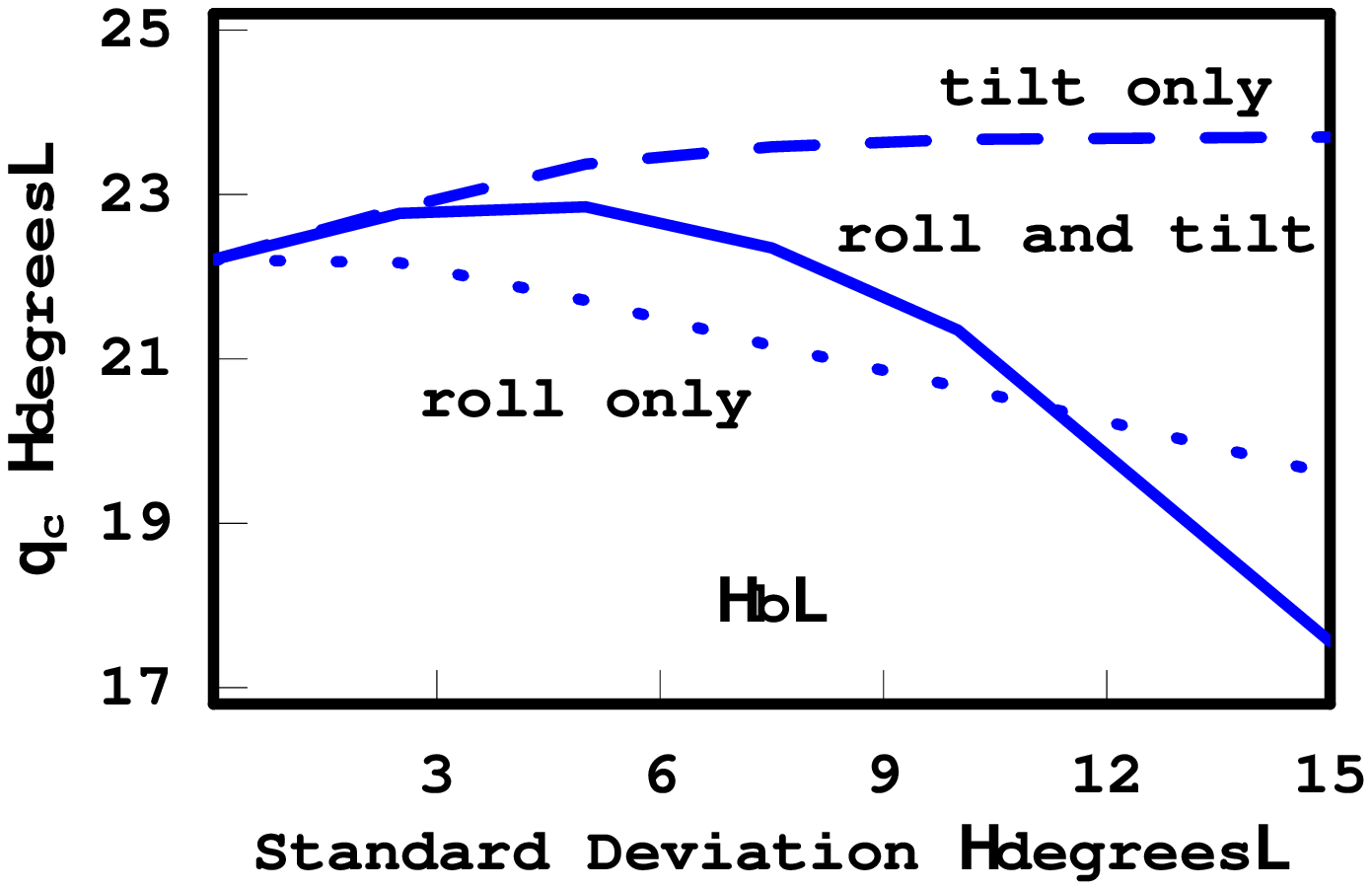}}
\caption{The dependence of $\theta_{c}$ for a monomer pile as 
surface properties are changed from a randomly oriented, flat, 
close-packed surface with no roll. 
(a) An increase in the normalized edge length $a$ for equilateral traps
stabilizes the traps and increases $\theta_c$. 
(b) Individual effects of including a Gaussian roll distribution 
(dotted line, destabilizing), tilt distribution (dashed line, stabilizing) 
and the combined effect of a simultaneous roll and tilt distribution with 
the same standard deviation (solid line, either stabilizing or destabilizing).
}
\label{figstability}
\end{figure}

Fig.~\ref{figstability} depicts how $\theta_c$ for a monomer pile 
changes as a function of change in (a) the trap size parameter $a$, 
and (b) the standard deviations of roll and tilt distributions,
both individually and jointly. From these plots, it becomes clear
that we need additional information about the grain-scale roughness
of the surface in order to quantitatively predict $\theta_c$ for
the monomer-dimer piles.

\end{multicols}

\begin{figure}
\centerline{\includegraphics[width=3in]{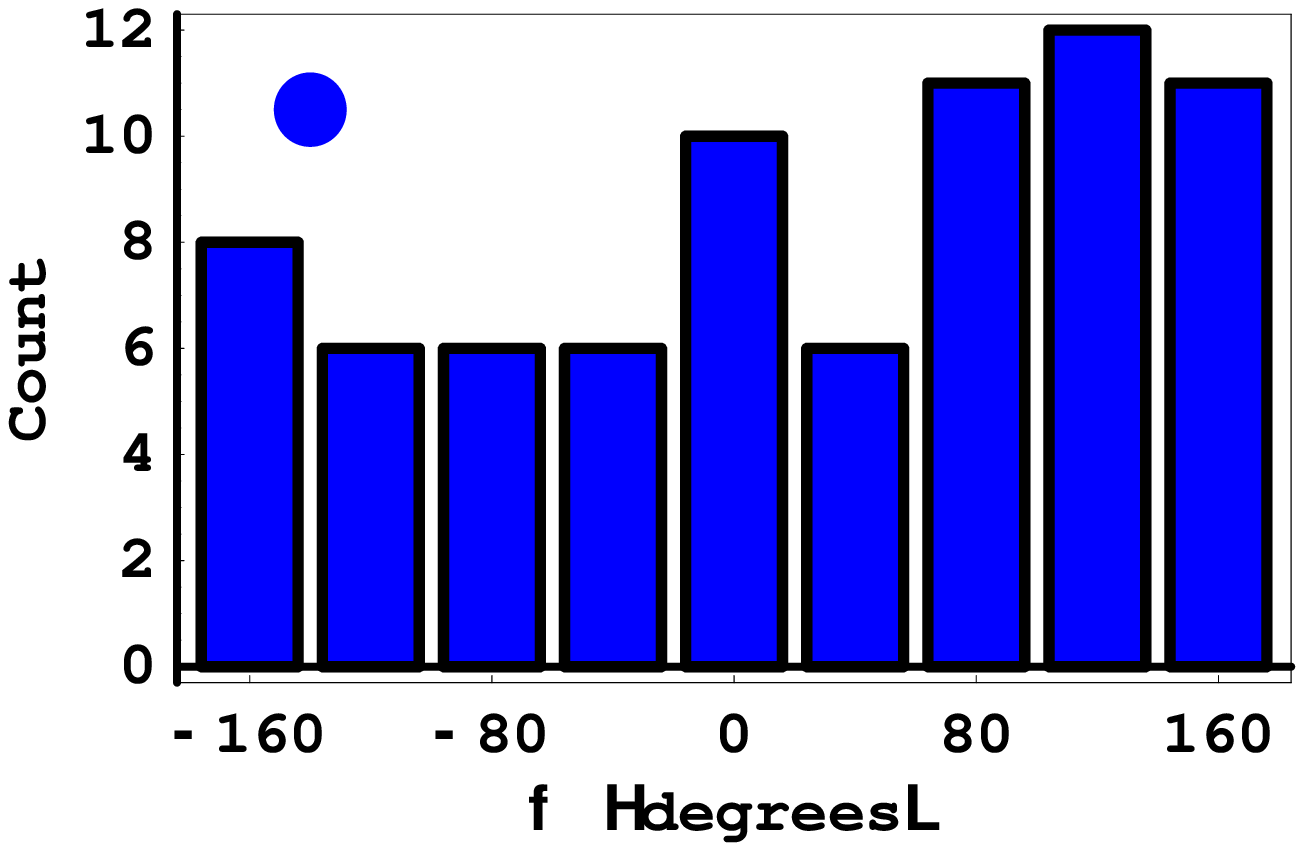}
\includegraphics[width=3in]{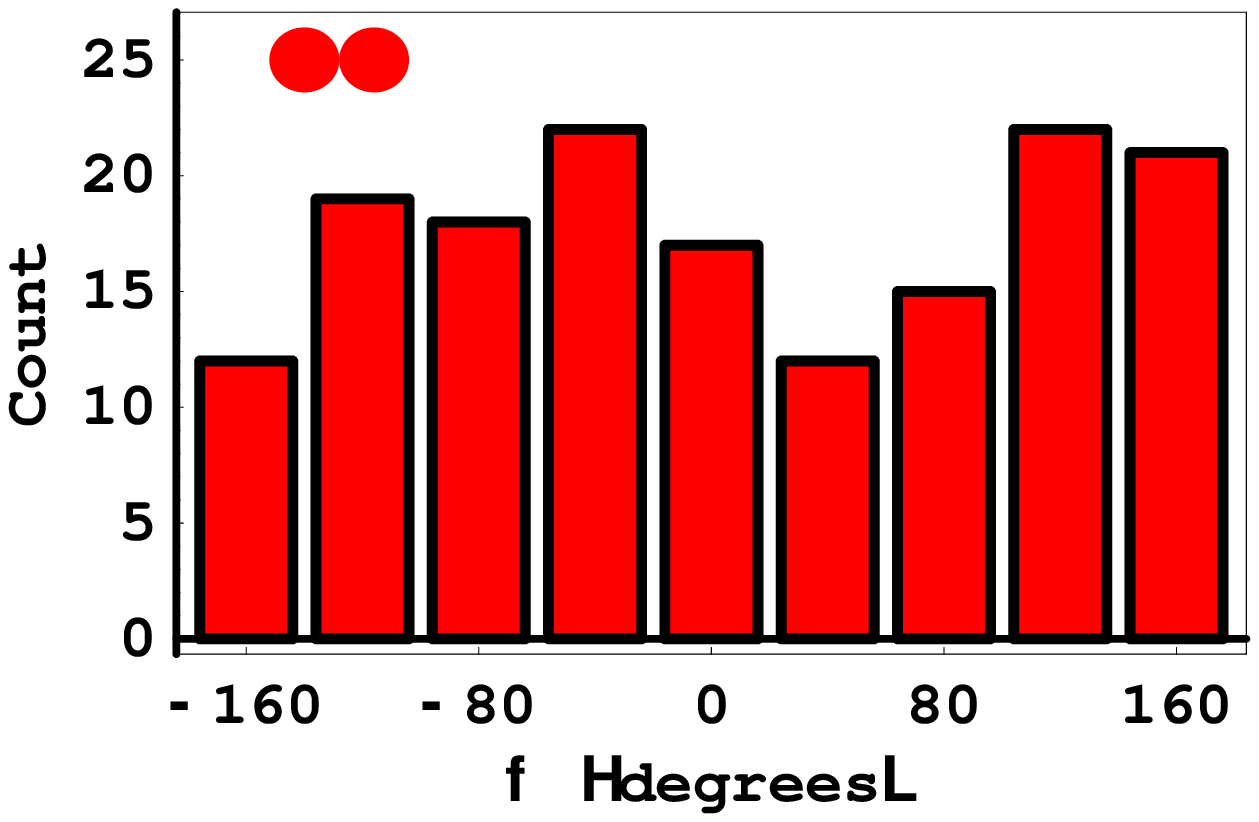}}
\centerline{\includegraphics[width=3in]{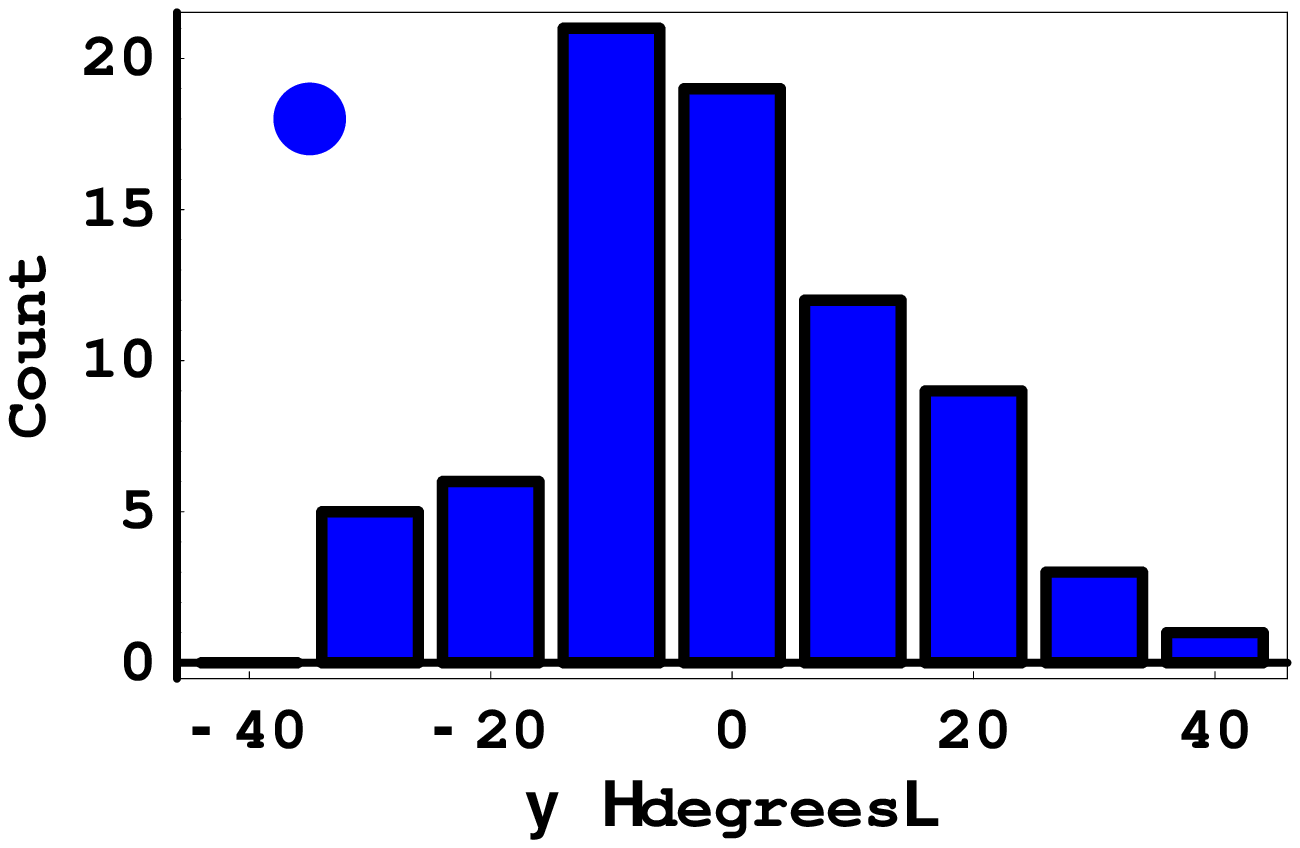}
\includegraphics[width=3in]{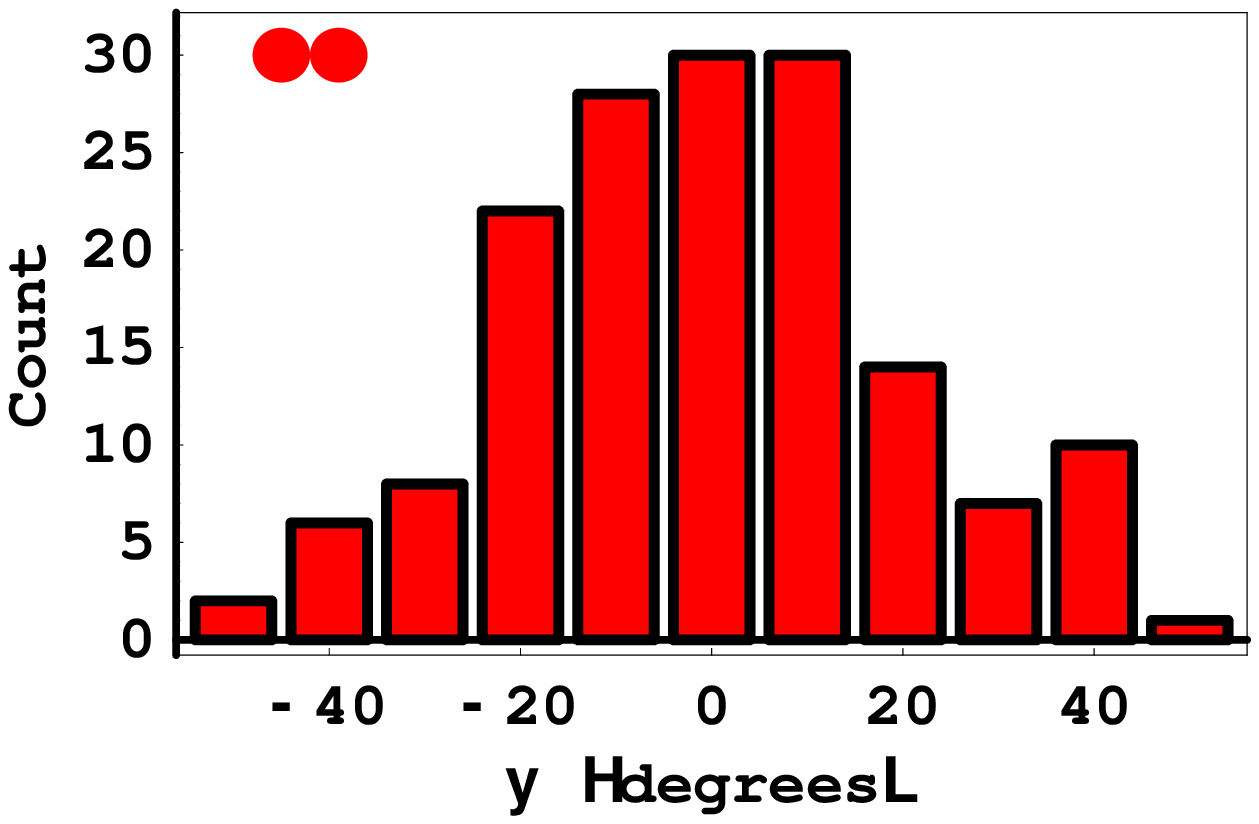}}
\centerline{\includegraphics[width=3in]{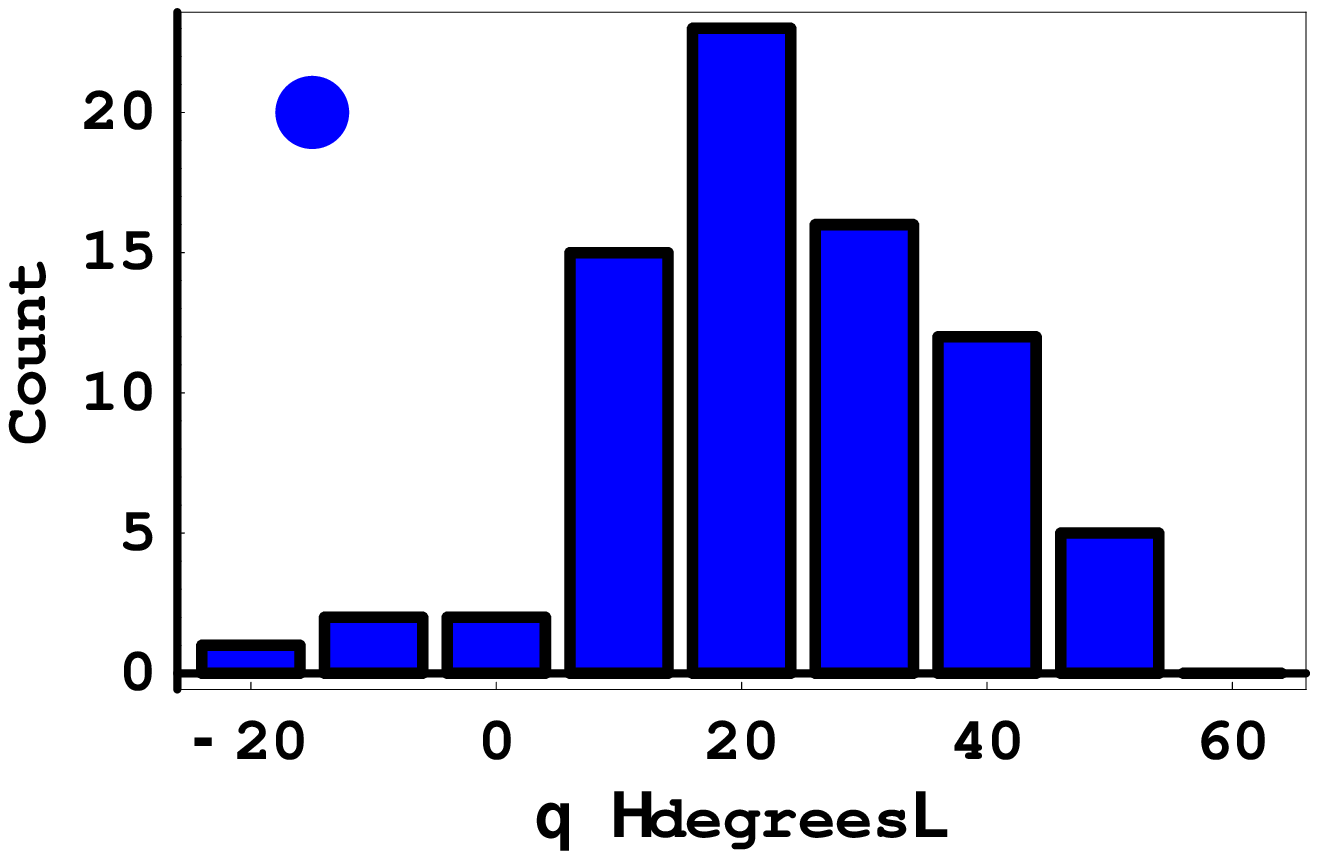}
\includegraphics[width=3in]{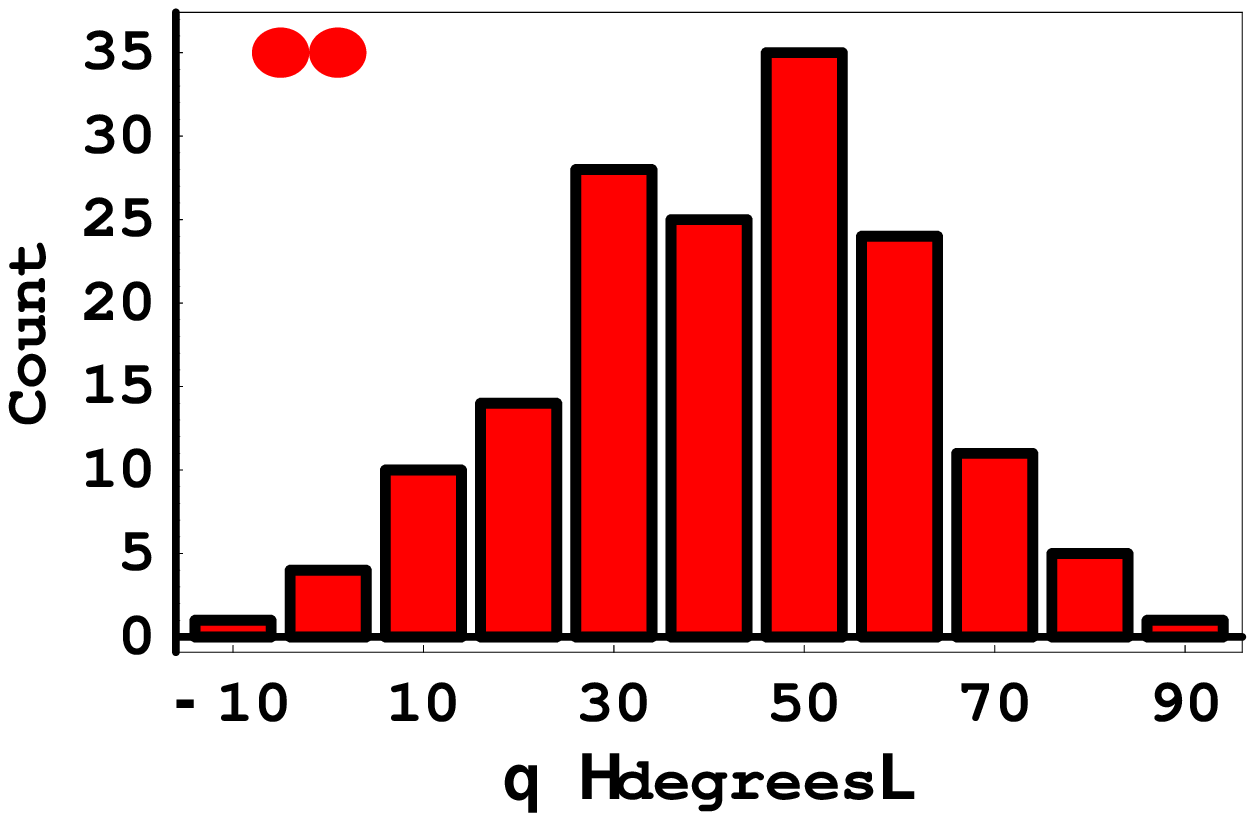}}
\caption{Histograms of yaw, roll and tilt distributions for the 
piles shown in Fig.~\protect\ref{figpiles}.
{\it Left:} monomer pile, {\it Right:} dimer pile.}
\label{fighist}
\end{figure}

\begin{multicols}{2}

In order to test whether real surfaces of piles exhibit the
assumed behavior, and to obtain representative values for the
average trap size and the width of yaw, roll and tilt distributions,
we imaged a portion of the surface of a monomer and a dimer 
pile (see Sec.~\ref{secexperiment}). The shape and orientation of
surface traps were identified as follows:
After locating the centers of the particles on the surface layer  
by stereographic imaging, we computed the average surface of the plane
by a least square fitting of the centers of mass to a plane.
We then performed a Delaunay triangulation of the particles projected
on to this plane in order to identify all base triangles
associated with potential surface traps. We then measured 
the yaw, roll and tilt of all the base triangles and created
histograms. We observed a uniform yaw distribution within a 
characteristic sampling error, justifying the use of 
Eq.(\ref{eqpyaw}). We also determined the standard deviations 
$\{\sigma_{\psi},\sigma_{\theta}\}$ for the roll and tilt histograms.
The histograms are shown in Figs.~\ref{fighist}
for the monomer and dimer pile. 

\begin{figure}
\centerline{\includegraphics[width=3in]{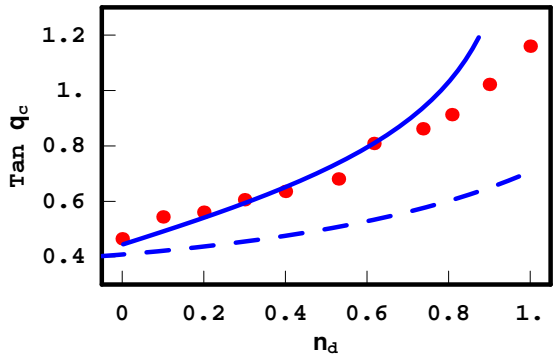}}
\caption{Measured and calculated values of $\tan \theta_c$ as a function
of dimer weight fraction $\nu_d$. Circles: Experiment. 
Solid Line: Computation based on the statistical description
of surface traps given by Eqs.(\protect\ref{eqps})-(\protect\ref{eqptilt}).
Dashed Line: Comparison to the earlier result based on a randomly oriented, 
flat, close-packed surface, reproduced from Fig.~\protect\ref{fignodistcomp}.}
\label{figthetac2}
\end{figure}

The comparison between the monomer and dimer piles revealed a moderate
increase of $\sigma_\psi$ and $\sigma_\theta$ from about $15^\circ$
to $20^\circ$, indicating a roughening of the surface along with 
the originally observed reduction in packing fraction. The average
edge length increased from $1.13$ to $1.18$, in agreement with 
Eq.~(\ref{eqa}). Due to the modest changes
in these parameters, we have used the trap characteristics obtained
from the monomer pile in the computation of $\theta_c$ for all the
mixture piles. Integrating the stability diagram shown in
Fig.~\ref{figthetamax}b with the PDFs given in Eqs.(\ref{eqps}-
\ref{eqptilt}) to obtain the appropriate $DOS(\theta)$ through a  
generalized form of Eq.(\ref{eqDOS}), we finally 
compute $\theta_c^{\rm mix}(\nu_d)$ through Eq.(\ref{eqfstab}). 
The result is shown in Fig.~\ref{figthetac2} as a solid line, and 
agrees well with experiment for $0<\nu_d<0.6$, particularly
considering that all the parameters have been provided by
independent measurement. No adjustable parameters remain 
in the model, suggesting that the assumption of monomer failure 
at the surface is valid in this range. The disagreement at larger 
$\nu_d$ is not surprising, given that the pile is comprised almost 
entirely of dimers and the assumption of monomer failure in the 
theory is expected to break down in this limit, resulting
in an over-estimate of the stability of the pile. 

\section{conclusion}
\label{secconc}

By introducing dimer grains in a sandpile comprised of rough spherical monomer 
grains, we have shown that the critical angle of the sandpile can be nearly 
doubled in the limit of high dimer content. Qualitatively, this result is not so 
surprising, given the significant number of previous measurements of the angle 
of repose of mixtures of cylindrical or spheri-cylindrical grains with spherical 
grains that have also shown an increase. However, the use of 
dimers, rather than other elongated objects, permits the random surface of the 
pile to be described as a collection of spheres that form triangular surface 
traps. These triangles have a distribution of edge lengths and yaw, roll, and 
tilt angles that can be directly obtained through stereo imaging and can be 
included in a theory. By comparing the macroscopic average property of the pile, 
the critical angle, with the grain-scale structure on the pile's surface 
obtained through imaging, we are able to show which aspects of the 
surface structure are important for determining the value of the 
critical angle.

For instance, the treatment of the critical angle of a random pile that 
considers only the mean angle of stability for a grain on 
a close packed surface, averaged over the yaw angle, may give a value close to 
the measured $\theta_c$, but is this just a fortuitous agreement? Our results 
show that, by averaging over realistic distributions of yaw and tilt, the more 
realistic median critical angle drops below the observed $\theta_c$. However, 
because the pile is random, the intergrain separation has a distribution itself, 
and the average edge length of a triangular surface trap is slightly greater 
than the grain diameter. This increase in the edge length of the surface trap, 
as compared to a perfectly close packed surface, increases the stability of a 
sphere in the trap. Indeed, we believe that it is the combination of the 
destabilizing influence of the roll distribution, along with the 
stabilizing influence of the larger edge length that gives the random pile of 
rough spherical grains its rather well-established value of 
$\theta_c \approx 23^\circ$.

These results for mixed monomer-dimer sandpiles shed some light on the observed 
initial increase in the critical angle of wet sandpiles, independent
of container size and liquid surface tension, when liquid content is below
a threshold value.\cite{Mason,Tegzes}. The initial linear increase in
$\tan\theta_c$ with $\nu_d$ may provide a plausible 
mechanism, in which the fraction of strongly wetted intergrain contacts 
increases gradually until all intergrain contacts are nearly uniformly wetted.
Provided that the formed bonds are strong enough and relatively dilute, 
the wet pile can be expected to respond similarly to a pile with a small 
fraction of dimers. One would expect that clusters with grains having more 
than one cohesive contact with neighboring grains would form in the wet 
sandpile as the threshold volume fraction is approached, and, above the 
threshold volume fraction, the picture of an average cohesive force 
holding grains together everywhere in the pile would become tenable.
The data in Refs.~\cite{Mason} and \cite{Tegzes} suggest an equivalent
$\nu_d$ of about 0.12 at the threshold liquid volume fraction.

It may be possible to extend the presented work to systems involving 
trimers and higher order clusters of grains\cite{Olson}. However,  
such clusters can have many different shapes, and to simplify the 
theoretical treatment, it may be necessary to restrict allowed shapes 
to close-packed or linear structures. Along a different direction, 
reducing the grain-grain friction coefficient $\mu$ will allow sliding 
failure modes and thereby lower the critical angle of the sandpile. 
Densification of the pile through tapping might also change the angles of 
stability. Finally, developing a theoretical understanding of the reduction 
of the grain packing fraction with increasing dimer content would help shed 
light on how strong intergrain attractive forces can alter the bulk 
structure of a random pile.

We thank P.~Chaikin, Z.~Cheng, G.~Grest and P.~Schiffer for stimulating 
discussions and suggestions. AJL was supported in part by the National
Science Foundation under Award DMR-9870785.

\begin{appendix}

\section*{Parameterization of a surface trap - yaw, roll and tilt}

In this Appendix, we define the parameters that describe 
the shape and orientation of a base triangle that connects the centers
of mass of the three supporting spheres that form a surface trap.

\begin{figure}
\centerline{\includegraphics[width=3in]{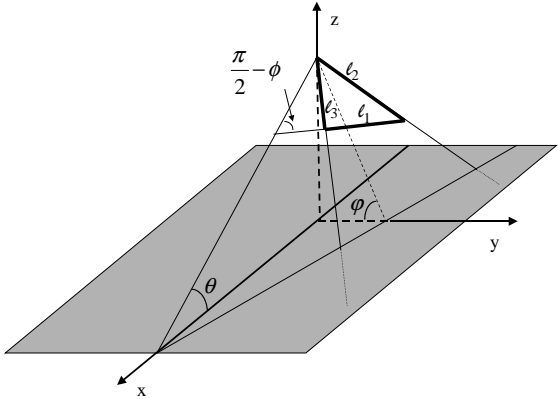}}
\caption{The geometry of a surface trap, characterized by its 
yaw $\phi$, roll $\psi$ and tilt $\theta$, as well as the edge 
lengths $\{l_i\}$ of its base triangle.}
\label{figgeom}
\end{figure}

The geometry is shown in Fig.~\ref{figgeom}. The coordinate
system is fixed such that gravity is in the $-z$ direction,
and the pile, whose mean surface is initially in the $xy-$plane, 
is ``tilted" by rotating it around the $x-$axis. 
Since overall translation of the triangle has no effect on 
trap stability, the vertex across the shortest edge (with length
$l_1$) has been arbitrarily placed on the $z-$axis for 
ease of illustration. The base triangle
can be fully specified by the positions of the two remaining
vertices relative to the first one; this leaves six parameters 
to be determined. 

A more useful parameterization than the relative positions 
of the vertices can be given as follows (See Fig.\ref{figgeom}):  
The shape of the base triangle is characterized by the lengths
of its edges. The two remaining edge lengths can be 
unambiguously labeled as $\l_2$ and $l_3$ anticlockwise around
the triangle when observed from a viewpoint above (at large $z$).
This leaves three angles that determine the 
orientation. The plane in which the base triangle resides 
is described by {\it roll} $\psi$ and {\it tilt} $\theta$. 
Similarly, the orientation of the triangle in the plane with 
respect to the downhill direction is described by the 
{\it yaw} $\phi$, as depicted in Fig.~\ref{figgeom}\cite{terminology}.

Given these parameters, the original base
triangle can be reconstructed (modulo translations) 
as follows: Place a triangle
with given edge lengths in the $xy-$plane such that the 
shortest edge is parallel to the $x-$axis and ``downhill" 
from the vertex across it, i.e., the $y-$ordinate of the 
vertex is larger.
Then, rotate the triangle around the $z-$ axis by $\phi$,
$y-$ axis by $\psi$ and finally, $x-$axis by $\theta$. 
With the proper labeling of vertices as described above,
every triangle is uniquely identified
except for degenerate cases (isosceles and equilateral 
triangles), in which case the stability criteria are 
identical and the particular choice of angles is immaterial.


This parameterization has two main advantages, both
of which facilitate statistical averaging over many traps,
performed in Sec.~\ref{secanalysis}:
 
(i) The shapes and orientations of surface traps are 
very likely to be statistically independent of each other,
and therefore they will have independent probability 
distributions. Splitting the parameters that describe 
these two attributes avoids dealing with joint probability 
distributions across these two classes of parameters.

(ii) Tilting the pile does not change the yaw and roll of a 
surface trap. Thus, a "stability interval" 
$[\theta_{min},\theta_{max}]$ can be defined for a surface 
trap of given yaw and roll, corresponding to all the values
of tilt $\theta$ for which the trap can stably support a 
surface particle.

\end{appendix}

\end{multicols}


\begin{references}
\bibitem{error}This value was erroneously
reported in Ref.~\protect\cite{Barabasi} as $23.4^\circ$.
\bibitem{commutation}The distributions of roll and tilt would not 
be identical since the two finite rotations do not commute. However,
the difference is only third order in the angle and quite small
for reasonably narrow distributions.
\bibitem{terminology}We have borrowed aviation/navigation
terminology to name the three angles, roll, yaw and tilt, that describe
the orientation of the base triangle.
\end{references}
\end{document}